 \documentclass[onecolumn,preprintnumbers,amsmath,amssymb]{revtex4}

\begin{document}


 \title{Bose-Einstein condensates in a homogeneous gravitational field}

 \author{J. I. Rivas}
 \email{jirs@xanum.uam.mx}\affiliation{Departamento de F\'{\i}sica,
 Universidad Aut\'onoma Metropolitana--Iztapalapa\\
 Apartado Postal 55--534, C.P. 09340, M\'exico, D.F., M\'exico.}

 \author{A. Camacho}
 \email{acq@xanum.uam.mx} \affiliation{Departamento de F\'{\i}sica,
 Universidad Aut\'onoma Metropolitana--Iztapalapa\\
 Apartado Postal 55--534, C.P. 09340, M\'exico, D.F., M\'exico.}

 \date{\today}

 \begin{abstract}
 The behavior of a Bose--Einstein condensate in a homogeneous gravitational field is
 analyzed. We consider two different trapping potentials. Firstly, the gas is inside a finite container.
 The effects of the finiteness of the height
 of the container in connection with the presence of a homogeneous gravitational
field are mathematically analyzed and the resulting energy
eigenvalues are deduced and used to obtain the corresponding
partition function and the ensuing thermodynamical properties.
Secondly, the trapping potential is an anisotropic harmonic
 oscillator and the effects of the gravitational field and of the
 zero--point energy on the condensation temperature are also
 considered. These results are employed in order to put forward an
 experiment which could test the so called Einstein Equivalence
 Principle.
 \end{abstract}

 \maketitle
\section{Introduction}

 Gravity can be understood at the classical level as a purely geometric effect, i.e., the
motion of a free classical particle moving in a curved manifold is
given by the Weak Equivalence Principle (WEP) \cite{Will}. In
other words, a particle moves along the geodesics of the
corresponding manifold. The introduction of additional
interactions is done resorting to the so--called Einstein
Equivalence Principle \cite{Will}, the famous ``semicolon goes to
coma rule''. This principle tells us that locally the laws of
physics are the special--relativistic laws, we may rephrase this
statement asserting that locally the gravitational field can be
gauged away

Additionally, we may state that the role of geometry is,
classically, local. This phrase means that the dynamics of a free
cla\-ssi\-cal particle located at a certain point $P$ of any
Riemannian manifold is, according to General Relativity (GR),
determined by the geometric properties of this manifold at $P$
(the motion equations can be written in terms of the connection
coefficients, which at $P$ depend only on the values of the
components of the metric and their derivatives evaluated at $P$),
geometry at any other point plays no role in the determination of
the motion when the particle is at $P$. If we consider the
geodesic deviation between two particles, then we would obtain
information of the Riemann tensor, but once again only of the
region where the motion of these classical particles takes place.

The experimental tests of GR contain a large number of proposals
and none of them contains an experimental output which could be
considered a counterexample to the theoretical predictions of GR
\cite{Will, Ciufolini}. This last statement could be misleading
since it could lead us to conclude that there are no conceptual
problems in connection with GR. Nevertheless, we may assert that
nowadays gravitational physics stands before a dead end. Indeed,
on one hand the dominating belief states that gravity shall be
described by a theory founded, among other concepts, upon quantum
ideas \cite{Rovelli, Kiefer, Zwiebach}. On the other hand, there
is, in this direction, no complete theory, neither mathematically
nor physically, in spite of the advocates of these ideas which
claim to be on the verge of deducing the final theory. These
aforementioned facts have spurred a large number of theoretical
attempts which can be considered in the phenomenological realm
\cite{Bluhm, Mattingly, Giovanni1, Giovanni2}. In other words, the
quest for experimental results, either looking for new effects
(for instance, the breakdown of Lorentz symmetry), or testing the
experimental postulates of general relativity, has acquire a
significative relevance. Recently the possibility of resorting to
the Fermi telescope and use it for the detection quantum gravity
effects stemming from light dispersion for high energy photons has
been put forward. A deeper and ampler analysis of this proposal
can be found in \cite{Giovanni3}.

In this context the present work addresses the behavior of a
Bose--Einstein condensate under the influence of a homogeneous
gravitational field. The idea is to consider the possibility of
testing the Einstein Equivalence Principle resorting to the
temperature as the parameter to be monitored. Indeed, the usual
Bose--Einstein experiments are carried out with the experimental
device at rest with respect to the Earth's surface
\cite{Pethick1}. We will deduce the changes in the condensation
temperature as a consequence of the presence of a homogeneous
gravitational field. This will be done for two different kind of
trapping potential, namely, a container of volume $V$ and an
anisotropic three--dimensional harmonic oscillator. It will become
clear that there is indeed a modification due to the presence of a
non--vanishing value of $g$, the acceleration of gravity. This
result is no surprise since we know that the condensation
temperature emerges at that point at which the chemical potential
equals the energy of the ground state \cite{Dalfovoro}. The
introduction of a non--vanishing homogeneous gravitational field
modifies the ground state of a particle either trapped by a
container of volume $V$ or by an anisotropic three--dimensional
harmonic oscillator. An interesting point in this context is
related to the fact that the change in the condensation
temperature, here we mean if it grows or decreases with respect to
the case in which $g=0$, depends upon the trapping potential.
Indeed, it will be shown that for a Bose gas trapped inside a
container of volume $V$ the condensation temperature, due to
$g\not =0$, is larger than for $g =0$, whereas, for the harmonic
oscillator case it is smaller.

Let us now explain a little bit deeper the whole idea. According
to the so--called Einstein Equivalence Principle (EEP) in a freely
falling frame, for sufficiently short experimental times, the
result of any experiment coincides with the outcome when $g=0$,
i.e., locally gravity can be gauged away \cite{Will}. The idea is
to perform the experiment in a freely falling frame and detect the
condensation temperature. If EEP is valid, then this temperature
shall be equal to the case in which $g=0$. Afterwards, knowing the
result of the same experiment when carried out on the Earth's
surface we may compare the corresponding experimental outputs can
contrasted them against the theoretical predictions arising from
EEP. In this sense we would have a test of EEP resorting to the
concept of temperature. Maxwell--Boltzmann (MB) statistics
predicts the appearance of condensation only if the temperature of
the system vanishes. Indeed, for a gas, obeying MB, comprising $N$
non--interacting particles and discarding also internal degrees of
freedom, the internal energy $U$ satisfies the condition $U = N<K>
= 3N\kappa T/2$, according to the theorem of equipartition of
energy \cite{Phatria}, here $<K>$ denotes the average kinetical
energy per particle, and $\kappa$ is Boltzmann constant. This last
expression tells us that $<K> =0$ if and only if $T=0$. The lowest
energy appears only when $T=0$. On the other hand, the case of
Bose--Einstein statistics is quite different. Indeed, this
situation involves a certain temperature, denoted condensation
temperature $T_c$, at which a large number of particles are in the
corresponding ground state, usually the behavior has the form
$N_0/N\sim (1-T/T_c)^{\alpha}$, where $N_0$ is the number of
particles in the ground state and $\alpha$ is a positive real
number which depends on the trapping potential, the number of
dimensions, etc \cite{Pethick1}. We may rephrase the previous
argumentations stating that classical behavior (here this
statement means MB) predicts condensation only if $T=0$, whereas,
quantum properties, Bose--Einstein statistics, entail a
macroscopical number of particles in the ground state even if
$T>0$. In this sense we could discuss the possible interpretation
of the present proposal as a quantum test for EEP since classical
statistics requires inexorably $T=0$.
\bigskip
\bigskip

\section{Condensation in a homogeneous gravitational field}
\bigskip

\subsection{Gas in a container}
\bigskip

As a first example let us consider the case of a finite container
immersed in a homogeneous gravitational field, for the sake of
simplicity we will assume that the gravitational field is along
the $z$--axis and that the container has cylindrical symmetry with
its axis parallel to $z$. The problem of a quantum particle in a
gravitational field is usually solve assuming that the particle
vanishes in the limit $z\rightarrow \infty$, i.e., the particle is
to be found in the interval $z\in [0, \infty]$. Under these
conditions the solution reduces to one of the so--called Airy
functions \cite{Fluegge}. Our situation is quite different,
indeed, the potential along the $z$--direction is comprised by two
terms, namely,

\[V(\textbf{z}) =
   \begin{cases}
         0, & \mbox{para $\textbf{0}<\textbf{z}<\textbf{L}$} \\
         \infty, & \mbox{elsewhere,}
   \end{cases}\]\\

whose eigenvalues and eigenfunctions read:

\begin{eqnarray}
E_{n}^{(0)}=\frac{{\hbar}^{2}{\pi}^{2}}{2m{L}^{2}}{n}^{2},
\label{polar1}
\end{eqnarray}

\begin{eqnarray}
\psi(z)=\sqrt{\frac{2}{L}}\sin\Big(\frac{n\pi
z}{L}\Big).\label{polar14}
\end{eqnarray}

Clearly, we must add an extra term containing the gravitational
interaction

\begin{eqnarray}
V_g = mgz. \label{polar2}
\end{eqnarray}

This term will be treated as a perturbation for the Hamiltonian
given by

\begin{eqnarray}
H_0= \frac{p^2_z}{2m} + V(\textbf{z}).
\label{polar3}
\end{eqnarray}

In other words, our complete Hamiltonian is

\begin{eqnarray}
H = H_0 + \lambda V_g. \label{polar5}
\end{eqnarray}

A lengthy calculation, up to second order in $\lambda$, renders
the following eigenenergies

\begin{eqnarray}
{\varepsilon}_{n}=\frac{{\hbar}^{2}{\pi}^{2}}{2m{L}^{2}}{n}^{2}+\lambda\frac{m
g L} {2}+{\lambda}^{2}\frac{4{m}^{3}{g}^{2}{L}^{4}}{{\hbar}^{2}
{\pi}^{6}}\Big(\frac{\zeta(5)}{n}-\frac{{\pi}^{4}}{48{n}^{2}}\Big).\label{polar4}
\end{eqnarray}

Of course, $n$ is an integer equal or larger than 1. This last
expression takes into account only the contribution to the energy
as a consequence of the Hamiltonian along the $z$ axis. It is
obvious that the contributions of the corresponding Hamiltonians
along $x$ and $y$ have to be included. These contributions are
simply the eigenenergies of a particle inside a two--dimensional
box whose form is a square (whose sides have a length equal to
$a$), namely,

\begin{eqnarray}
{\varepsilon}_{(n_x, n_y)}=
\frac{{\hbar}^{2}{\pi}^{2}}{2m{a}^{2}}\Big({n_x}^{2} +
{n_y}^{2}\Big).\label{polar5}
\end{eqnarray}

The energies are provided by the sum of these last two
expressions.

Statistical Mechanics teaches us that the knowledge of the
associated eigenenergies allows us to calculate the corresponding
partition function \cite{Phatria}. In our case we will resort to
the Grand--Canonical partition function, the one tells us that the
number of particles in thermodynamical equilibrium is given by

\begin{eqnarray}
N=\sum_{\varepsilon}\langle{n}_{\varepsilon}\rangle=\sum_{\varepsilon}\frac{1}{{z}^{-1}{e}^{\beta\varepsilon}-1}.
\label{polar5}
\end{eqnarray}

Here $\beta=\frac{1}{\kappa T}$ and $z$ is the so--called
fugacity, defined as follows: $z={e}^{\frac{\mu}{\kappa T}}$, with
$\mu$ the chemical potential. This last expression can be cast in
the following form

\begin{eqnarray}
N=\frac{z}{1-z}+\sum_{{\varepsilon}_{A}=1}^{\infty}\sum_{{\varepsilon}_{w}=1}^{\infty}\frac{1}{{z}^{-1}{e}^
{\beta[{\varepsilon}_{A}+{\varepsilon}_{w}]}-1}. \label{polar6}
\end{eqnarray}

In this last expression it has been introduced the fact that the
energy can be divided into two contributions, namely; (i) one
stemming from the motion on the plane perpendicular to the
gravitational field, this term has been denoted by
$\varepsilon_{A}$; (ii) the energy coming from the motion along
the $z$--axis, $\varepsilon_{w}$.

Particles in the ground state are given by

\begin{eqnarray}
{N}_{0}=\frac{z}{1-z}. \label{polar7}
\end{eqnarray}

The number of particles in the excited states is given by $N-N_0$,
which for our particular case takes the form

\begin{eqnarray}
 N-{N}_{0}=\frac{V}{{\Lambda}^{3}}{g}_{3/2}(z{e}^{-\frac{m g L}{2 \kappa T}})
 . \label{polar8}
\end{eqnarray}

Here ${g}_{\nu}(z)$ are called Bose--Einstein functions and
$\Lambda=\frac{{(2\pi m \kappa T)}^{\frac{1}{2}}}{h}$ is de
Broglie thermal wavelength \cite{Phatria}.

We may at this point understand that the presence of a
gravitational field does indeed impinge upon the condensation
temperature, among other thermodynamical parameters. To fathom
this last statement, let us recall that condensation appears (when
it is possible) at that temperature in which the chemical
potential equals the ground energy. For our particular case this
last assertion entails

\begin{eqnarray}
 \mu(T = T_c) = \frac{m g L}{2}. \label{polar9}
\end{eqnarray}

This last expression can be contemplated as an implicit definition
of the condensation temperature, $T_c$. Indeed, for this case we
obtain

\begin{eqnarray}
{({T}_{c}^{o})}^{3/2}={T}_{c}^{3/2}\Big[1-\frac{1}{\zeta(3/2)}
\sqrt{\frac{\pi m g L}{2 \kappa {T}_{c}}}\Big].\label{polar10}
\end{eqnarray}

This last expression can be approximately written as follows

\begin{eqnarray}
 {T}_{c}\approx {T}_{c}^{o}\Big[1 + \frac{2}{3} \frac{1}{\zeta(3/2)}\sqrt{\frac{\pi m g L}{2
 \kappa {T}_{c}^{o}}}\;\Big],\label{polar11}
\end{eqnarray}

Here we have introduced the following definition which is the
condensation temperature without gravitational field.

\begin{eqnarray}
{T}_{c}^{o}=\frac{{h}^{2}}{2\pi m
\kappa}{\Big(\frac{N}{V\zeta(3/2)}\Big)}^{2/3}.\label{polar12}
\end{eqnarray}

Our last results show clearly that there is an increase in the
condensation temperature provoked by the presence of a
non--vanishing homogeneous gravitational field. Since the
achievement of very low temperatures is an experimental difficulty
this fact seems to be an advantage. Nevertheless, there are two
points that have to be underlined in connection with this first
example. Firstly, current technology achieves condensation
resorting to trapping potential which are not our simple box.
Secondly, this feature, increase of $T_c$ due to the presence of a
non--vanishing $g$, is not a general  behavior, as will be shown
below.
\bigskip

\subsection{Gas in a harmonic trap}
\bigskip

As mention before, in the experimental realm the condensation
process does not resort to a gas within a container, the trapping
potential has a more sophisticated structure. Indeed, there are
several kind of traps, for instance magneto--optical traps (MOT),
Optical traps (OT), etc. \cite{Pethick1}. The mathematical
description of the available magnetic traps, at least for alkali
atoms, is that the corresponding confining potential can be
approximated by a three--dimensional harmonic oscillator

\begin{equation}
U(x,y,z)=\frac{m}{2}\Big({w}_{1}^{2}{x}^{2}+{w}_{2}^{2}{y}^{2}+{w}_{3}^{2}{z}^{2}\Big)
.\label{polar13}
\end{equation}

We now consider the presence of a homogeneous gravitational field
along the $z$--axis, hence, the complete potential becomes

\begin{equation}
U(x,y,z)=\frac{m}{2}\Big({w}_{1}^{2}{x}^{2}+{w}_{2}^{2}{y}^{2}+{w}_{3}^{2}{\Big(z+\frac{g}{{w}_{3}^{2}}\Big)}^{2}\Big)-
\frac{1}{2}\frac{m{g}^{2}}{{w}_{3}^{2}}.\label{polar14}
\end{equation}

The energy eigenvalues are given by

\begin{equation}
\varepsilon=\hslash{w}_{1}\Big({n}_{x}+\frac{1}{2}\Big)+\hslash{w}_{2}\Big({n}_{y}+\frac{1}{2}\Big)+
\hslash{w}_{3}\Big({n}_{z}+\frac{1}{2}\Big)-\frac{1}{2}\frac{m{g}^{2}}{{w}_{3}^{2}};\quad
{n}_{x},{n}_{y},{n}_{z}\epsilon\mathbb{N}.\label{polar15}
\end{equation}

The density of states is provided by

\begin{equation}
\Omega(\varepsilon)=\frac{{\Big(\varepsilon
+\frac{mg}{2{w}_{3}^{2}}-\frac{\hslash}{2}({w}_{1}+{w}_{2}+{w}_{3})
\Big)}^{2}}{2{\hslash}^{3}{w}_{1}{w}_{2}{w}_{3}}.\label{polar16}
\end{equation}

This last expression allows us to calculate the average number of
particles $N$

 \begin{equation}N=\int_{0}^{\infty}\frac{\Omega(\varepsilon)}{{z}^{-1}{e}^{\beta
\varepsilon}-1}d\varepsilon+\frac{1}{{z}^{-1}{e}^{\beta
{\varepsilon}_{0}}-1}. \label{polar17}
\end{equation}

Here the energy of the ground state reads

\begin{equation}
{\varepsilon}_{0}=\frac{\hslash}{2}({w}_{1}+{w}_{2}+{w}_{3})-\frac{1}{2}\frac{m{g}^{2}}
{{w}_{3}^{2}}.\label{polar17}
\end{equation}

It is readily seen that the presence of a non--vanishing
homogeneous gravitational field modifies the ground state energy,
and therefore, this field entails a change in the condensation
temperature. Indeed, it can be shown that a modification in the
ground state energy due to the presence of an interaction, as the
case here, does imply a new condensation temperature provided by
\cite{Pethick1}

\begin{equation}
\Delta T_c = -\frac{\zeta(2)}{3\zeta(3)}\frac{\Delta\mu}{\kappa}
.\label{polar28}
\end{equation}

For our particular case the change in the chemical potential,
$\Delta\mu = -(0.456)\frac{mg^2}{2\omega^2_z}$, namely, the
condensation temperature under the presence of a homogeneous
gravitational field reads

\begin{equation}
T^{(g)}_c =  T^{(0)}_c
-(0.456)\frac{mg^2}{2\kappa\omega^2_z}.\label{polar18}
\end{equation}

In this last expression $T^{(g)}_c$ is the condensation
temperature if theres is a non--vanishing gravitational field ,
whereas $T^{(0)}_c$ denotes the condensation temperature without
gravitational field. If $g=0$ we recover the usual condensation
temperature.

\subsection{Conclusions}
\bigskip

In the present work the effects upon the condensation temperature
of a homogeneous gravitational field have been calculated. This
has been done for two particular trapping potential, namely, a
container of volume $V$ and a three--dimensional harmonic
oscillator, the latter is a case which mimics some of the current
technological possibilities in the experimental realm.

It has been shown that for the case of a trapping potential
modelled by a container of volume V the presence of a homogeneous
gravitational field entails an increase of the condensation
temperature. For a three--dimensional harmonic oscillator the
presence of this simple gravitational field has precisely the
opposite effect, i.e., the condensation temperature suffers a
decrease. This last situation entails that, according to EEP, the
temperature of a freely falling condensate, trapped by a harmonic
oscillator, should be higher than the corresponding temperature if
the condensate lies at rest with respect to the surface of the
Earth.

In order to have a quantitative idea of the modifications upon the
condensation temperature due to the presence of a non--vanishing
homogeneous gravitational field, for the case of a gas trapped by
harmonic oscillators, we now show a table for three elements,
namely, Rubidium, Sodium, and Lithium. The frequency $\omega_z$ is
equal to $10^3 s^{-1}$, and the condensation temperatures (under
the presence of a gravitational field, i. e., $T^{(g)}_{c}$) read
$50\times10^{-9}$ K, $2\times10^{-6}$ K, and $300\times10^{-9}$ K,
for Rubidium, Sodium, and Lithium, respectively, \cite{Cornell,
Davis, Bradley}.

\begin{center}
\begin{tabular}{|c|c|} \hline
               Element    &Gas trapped by oscillators                  \\
                                                         &                                            \\
                           &$\frac{{T}^{(0)}_{c}-{T}^{(g)}_{c}}{T^{(g)}_{c}}$     \\ \hline
               ${}^{87}Rb$                                 &$0.0044$                                    \\
               ${}^{23}Na$                                 &$0.0031$                                    \\
               ${}^{7}Li$                                  &$0.0062$                                    \\ \hline
\end{tabular}
\end{center}

These last results display clearly the fact that, experimentally,
the best option seems to be Lithium since it has the largest
percentage of change, whereas  Sodium is the worst one. The
percentages are all smaller than one percent and oscillate between
$0.31$ and $0.62$.

\begin{acknowledgments}
JIRS acknowledges CONACyT grant No. 160453.
 \end{acknowledgments}

 \end{document}